\documentclass[12pt]{article}
\usepackage{latexsym}
\usepackage{amssymb,amsfonts,amsmath}
\usepackage{graphicx} 
\usepackage{indentfirst}
\usepackage{bbm}
\usepackage{amssymb}
\usepackage{verbatim}
\usepackage{amsmath, amsthm,amssymb}
\usepackage{mathrsfs}
\usepackage{hyperref}
\usepackage{amsfonts}
\usepackage{dsfont}
\usepackage{cite}

\newcommand {\cD}{{\cal D}}
\newcommand {\cE}{{\cal E}}

\newcommand {\cG}{{\cal G}}
\newcommand {\cH}{{\cal H}}

\newcommand {\cL}{{\cal L}}

\newcommand {\cN}{{\cal N}}

\newcommand {\cR}{{\cal R}}

\newcommand {\cT}{{\cal T}}

\newcommand {\cW}{{\cal W}}

\def\a{\alpha}
\def\b{\beta}
\def\c{\chi}
\def\d{\delta}

\def\f{\phi}

\def\l{\lambda}

\def\q{\theta}
\def\r{\rho}
\def\s{\sigma}

\def\x{\xi}

\def\J{\Psi}

\def\U{\Upsilon}

\def\re{{\rm e}}

\newcommand{\ad}{{\dot{\alpha}}}                           
\newcommand{\bd}{{\dot{\beta}}}                            

\newcommand{\pa}{\partial}                           
\newcommand{\hf}{\frac12}

\newcommand{\be}{\begin{equation}}
\newcommand{\ee}{\end{equation}}
\newcommand{\bea}{\begin{eqnarray}}
\newcommand{\eea}{\end{eqnarray}}
\newcommand{\non}{\nonumber}
\newcommand{\ba}{\begin{array}}
\newcommand{\ea}{\end{array}}

\def\double #1{#1{\hbox{\kern-2pt $#1$}}}

\newcommand{\sD}{\mathsf{D}}

\newcommand{\bsubeq}{\begin{subequations}}
\newcommand{\esubeq}{\end{subequations}}

\newcommand{\rd}{\mathrm d}
\numberwithin{equation}{section}

\begin{document}

\begin{center}
{\Large \bf 
Generalised Fayet-Iliopoulos terms in supergravity}
\end{center}

\begin{center}
{\bf Sergei M. Kuzenko} \\
\vspace{5mm}

\footnotesize{
{\it Department of Physics M013, The University of Western Australia\\
35 Stirling Highway, Perth W.A. 6009, Australia}}  
~\\
\vspace{2mm}
\end{center}

\begin{abstract}
\baselineskip=14pt
The U(1) vector multiplet theory with the Fayet-Iliopoulos (FI) term is one of the oldest and simplest models for spontaneously broken rigid supersymmetry. Lifting the FI term to supergravity requires gauged $R$-symmetry, as was first demonstrated in 1977 by Freedman within ${\cal N} =1$ supergravity. There exists an alternative to the standard FI mechanism,  which is reviewed in this conference paper. It is obtained by replacing the FI model with a manifestly gauge-invariant action such that its functional form is determined by two arbitrary real functions of a single complex variable. One of these functions generates a superconformal kinetic term for the vector multiplet, while the other yields a generalised FI term. Coupling such a vector multiplet model  to supergravity does not require gauging of the  $R$-symmetry. These generalised FI terms are consistently defined for any off-shell formulation for ${\cal N}=1$ supergravity, and are compatible with a supersymmetric cosmological term.
\end{abstract}

\renewcommand{\thefootnote}{\arabic{footnote}}


\section{Introduction}

\noindent
As is well-known, the observed accelerating expansion of the universe can be accounted for  by a small positive cosmological constant. It is therefore desirable to look for theoretical mechanisms that naturally lead to the required features 
of the cosmological constant, which are: (i) its positivity; and (ii) its small value.
In the last five years it has been widely appreciated 
that such a mechanism is potentially  provided by spontaneously broken local supersymmetry. In particular, this is one of the ideas 
which have been advanced within the framework of  pure de Sitter $\cN=1$ supergravity models, see e.g. \cite{ADFS,DFKS,BFKVP,HY,K15,AM,BMST}. 
Actually the observation of a connection between spontaneously broken local supersymmetry 
and a positive cosmological constant
is not new and, in fact, goes back to two seminal works published 
in 1977 \cite{Freedman,DZ}. In one of them, Freedman \cite{Freedman} derived the locally supersymmetric extension of the FI term by gauging the $R$-symmetry. 
Eliminating the auxiliary field $\sD$ yields a cosmological term  
with the cosmological constant being positive and 
proportional to $\x^2$, where $\x$ is the parameter appearing in 
the rigid supersymmetric FI action \cite{FI}
\bea
S=\frac{1}{2}  \int \rd^4 x \rd^2 \q   \,  W^\a W_\a 
-2\x  \int \rd^4 x \rd^2 \q  \rd^2 \bar{\q} \,V~,\qquad W_\a = -\frac 14 \bar D^2 D_\a V~,
\eea
with $W_\a$ the chiral gauge-invariant field strength \cite{FZ}. 
In the other work, Deser and Zumino  \cite{DZ}
elaborated on the super-Higgs effect \cite{VS} and demonstrated that the spontaneous breaking of local supersymmetry is accompanied by the appearance of 
a positive contribution to the cosmological constant proportional to the square of the scale of supersymmetry breaking.
The complete cosmological constant in \cite{DZ} also includes a negative
contribution coming from the supersymmetric cosmological term discovered by Townsend \cite{Townsend}. If a U(1) vector multiplet theory with the FI term is consistently coupled to supergravity, no supersymmetric cosmological term is permissible \cite{SW,BFNS},  as in the case of \cite{Freedman}.

It should be mentioned that both 1977 papers \cite{Freedman,DZ} discussed above made use of $\cN=1$ supergravity without auxiliary fields -- off-shell supergravity 
simply did not exist at the time. However, within a year the main ideas and results of  \cite{Freedman,DZ} were extended to off-shell supergravity. 
In the framework of the old minimal formulation for $\cN=1$ supergravity \cite{WZ,old1,old2},
the modern description of the FI term was given by Stelle and West \cite{SW}.\footnote{For the new minimal formulation \cite{new} the FI term was described in  \cite{SohniusW3}.} 
The supersymmetric cosmological term in old minimal supergravity was constructed in \cite{FvN2,Siegel}. A {\it complete} off-shell extension of the construction 
attempted\footnote{Since Deser 
and Zumino \cite{DZ} made use of supergravity without auxiliary fields, 
it was next to impossible to construct a 
complete supergravity-Goldstino action in their setting.}  
by Deser and Zumino \cite{DZ} was given by Lindstr\"om and Ro\v{c}ek \cite{LR} 
who proposed the first off-shell model for pure de Sitter supergravity
in four dimensions. 
They coupled a nilpotent covariantly chiral scalar 
to old minimal supergravity, with a supersymmetric cosmological term included. 
They reduced the theory to components and computed the same cosmological term
as in \cite{DZ} and also in \cite{BFKVP,HY,K15,AM,BMST}.

We see that some of the  results obtained in recent publications
 \cite{ADFS,DFKS,BFKVP,HY,K15,AM,BMST} were derived for the first time in the 
 late 1970s. Unfortunately, back in the 1970s nobody was interested in 
generating a positive cosmological constant. Everyone wanted it to vanish.

The FI terms can consistently be defined only for a restricted class of supergravity-matter theories. It is appropriate here to include a quote from the work \cite{BFNS} which identified 
such dynamical systems:  ``in  order for a U(1) gauge theory with a Fayet-Iliopoulos term to be consistently coupled to supergravity, preserving gauge invariance, the superpotential must be R invariant. 
A supersymmetric cosmological term and therefore an explicit mass-like term 
for the gravitino is forbidden by gauge invariance.'' It can be shown \cite{FGKV}
that all such theories can be realised within the new minimal auxiliary field formulation 
(and hence they possess dual realisations in old minimal supergravity).
Since these supergravity-matter systems are not extremely attractive for phenomenological applications, one might ask the following question:  
Could there exist a locally supersymmetric generalisation of, or an alternative to, 
the standard FI term that would be free of the limitations of the latter? 
This important question was posed for the first time shortly before Christmas 2017 
by Cribiori, Farakos, Tournoy and  Van Proeyen \cite{CFTVP} who 
also provided a positive answer by explicitly constructing 
a generalised FI term in supergravity. A whole family of generalised FI terms in supergravity, 
including the one constructed in \cite{CFTVP}, were proposed 
in \cite{K18,K19}.\footnote{Refs. \cite{CFTVP}  and \cite{K18} appeared in the arXiv within less than a month of each other.}
This  conference paper is a brief review of the results obtained in \cite{K18,K19}.


\section{The model} 

The constructions presented in \cite{K18,K19} are based on the assumption 
that local supersymmetry is in a spontaneously broken phase {\it ab initio}. 
In terms of a massless vector multiplet described by a gauge prepotential $V$,
this assumption is simply the requirement that the real gauge-invariant descendant 
$\cD \cW \equiv \cD^\a \cW_\a =\bar \cD_\ad \bar \cW^\ad  = \bar \cD \bar \cW$ 
is nowhere vanishing, i.e. $(\cD W)^{-1}$ exists. 
As usual,  $\cW_\a$ denotes the covariantly chiral field strength \cite{WZ}
\bea
\cW_\a := -\frac{1}{4} (\bar \cD^2 - 4R) \cD_\a V~,\qquad \bar \cD_\bd \cW_\a=0~,
\label{W}
\eea
which is invariant under gauge transformations of the form
\bea
\d_\l V = \l +\bar \l~, \qquad \bar \cD_\ad \l =0~.
\label{gaugefreedom}
\eea
The gauge prepotential is chosen to be super-Weyl inert, $\d_\s V=0$.\footnote{All relevant information concerning the supergravity covariant derivatives  $\cD_A =(\cD_a, \cD_\a, \bar \cD^\ad)$ and the super-Weyl transformations (with the super-Weyl parameter $\s$ being chiral, $\bar \cD_\ad \s=0$)
can be found in \cite{K19}.  }

In the phase with unbroken supersymmetry, there exists a unique 
gauge-invariant action for the vector multiplet coupled to conformal supergravity \cite{WZ}
\bea
S_{\rm Maxwell} [V] 
= \frac{1}{8}\int \rd^4 x \rd^2 \q  \rd^2 \bar{\q} \, E\,
 V \cD^\a (\bar \cD^2 -4R ) \cD_\a V
 =\frac{1}{2}  \int \rd^4 x \rd^2 \q   \, \cE\, \cW^2~, 
\eea
where $\cW^2 := \cW^\a \cW_\a$ and
$\cE$ is the chiral integration measure \cite{Siegel}.
In the spontaneously broken phase, however,  a more general
action can be introduced to describe the dynamics of 
a self-interacting vector multiplet coupled to conformal supergravity  \cite{K19}.
It has the form 
\bea
S [V] =
\frac{1}{2}  \int \rd^4 x \rd^2 \q   \, \cE\, \cW^2 +
 \int \rd^4 x \rd^2 \q  \rd^2 \bar{\q} \, E\,
\frac{\cW^2\,{\bar \cW}^2}{(\cD \cW)^2}\,
\cH \left(- \frac{\cD^2 \cW^2  }{(\cD \cW)^{2}}
\,, \, -\frac{ \bar \cD^2 \bar \cW^2 }{(\cD \cW)^{2}} \right)~,~~
\label{action}
\eea
where $\cH (z, \bar z)$ is a real function of one complex variable.

We now turn to the generalised FI term introduced in  \cite{K19}. It is given by 
\bea
{\mathbb J}^{(\cG)}_{\rm FI} [V; \U]=   \int \rd^4 x \rd^2 \q  \rd^2 \bar{\q} \, E\,
\U {\mathbb V}^{(\cG)}
~, \qquad {\mathbb V}^{(\cG)}:=
 {\mathbb V} \,\cG \left(- \frac{\cD^2 \cW^2  }{(\cD \cW)^{2}}
\,, \, -\frac{ \bar \cD^2 \bar \cW^2 }{(\cD \cW)^{2}} \right)~,
\label{generalised}
\eea
where $\mathbb V$ denotes the following composite gauge-invariant
scalar \cite{KMcAT-M} 
\bea
{\mathbb V} := - 4 \frac{\cW^2 \bar \cW^2}{(\cD \cW)^3}~, 
\eea
and
$\cG (z, \bar z)$ is a real function of one complex variable, which is  subject only 
to the  condition $ \cG(1,1)\neq0$ (see below) 
and {\it may} also depend on super-Weyl invariant matter superfields. 
In the right-hand side of \eqref{generalised} $\U$ is a real scalar with  super-Weyl transformation
\bea
\d_\s \U = (\s +\bar \s) \U~.
\label{2.7}
\eea
It is the nowhere vanishing scalar $\U$ which encodes information about a specific off-shell supergravity theory.
Within the new minimal formulation for $\cN=1$ supergravity
\cite{new,SohniusW3}, $\U$ can be identified with the corresponding 
linear compensator $ L$, 
\bea  
(\bar \cD^2 -4R) { L}  =0~, \qquad \bar { L}= { L}~.
\eea
In pure old minimal  supergravity
\cite{WZ,old1,old2}, $\U$ is given by 
$\U = \bar S_0 S_0$,
where  $S_0$ is 
the chiral compensator$, \bar \cD_\ad S_0 =0$,
with super-Weyl transformation law 
$\d_\s S_0 = \s S_0$. In the presence of chiral matter, however, $\U$ must be deformed, 
see below. 
In principle, the super-Weyl transformation law \eqref{2.7} is the only condition on $\U$.
This means that the generalised FI term \eqref{generalised} 
is consistently defined for any off-shell formulation for $\cN=1$ supergravity, 
unlike the standard FI term in supergravity.

We denote by ${\mathbb J}^{[n]}_{\rm FI} [V; \U]$ the
generalised FI term \eqref{generalised} corresponding to 
the homogeneous function $\cG(z,\bar z) = (z\bar z )^n$, with $n$ a real parameter. 
This one-parameter family of generalised FI terms was introduced in \cite{K18}. 
Its special representative ${\mathbb J}^{[-1]}_{\rm FI} [V; \U]$ was proposed for the first time
in  \cite{CFTVP}.

The composite superfield ${\mathbb V}^{(\cG)}$ defined in \eqref{generalised}
has three important properties. 
Firstly, it satisfies the nilpotency conditions
\bea
{\mathbb V}^{(\cG)}{\mathbb V}^{(\cG)}=0~, 
\qquad
{\mathbb V}^{(\cG)} \cD_A \cD_B {\mathbb V}^{(\cG)} =0~,
\qquad
{\mathbb V}^{(\cG)} \cD_A \cD_B \cD_C {\mathbb V}^{(\cG)}=0~,
\label{5}
\eea
which are characteristic of the Goldstino superfields studied 
in \cite{KMcAT-M,BHKMS}. 
Secondly, ${\mathbb V}^{(\cG)} $ is gauge invariant, $\d_\l {\mathbb V}^{(\cG)}=0$. 
Thirdly, 
${\mathbb V}^{(\cG)} $ is super-Weyl inert, $\d_\s {\mathbb V}^{(\cG)} =0$.  

The vector multiplet action \eqref{action}  
and the generalised FI term \eqref{generalised} are two sectors 
of the complete supergravity-matter action
\bea
S= S_{\rm SUGRA} + S[V] -2\x {\mathbb J}^{(\cG)}_{\rm FI} [V; \U] ~,
\label{CompleteAction}
\eea
where $S_{\rm SUGRA} $ denotes an action for supergravity coupled to other 
matter supermultiplets, for instance
\bea
S_{\rm SUGRA} &=& - 3
\int {\rm d}^{4} x \rd^2\q\rd^2\bar\q\,
E\,  \bar S_0  \,
\re^{-\frac{1}{3} K(\f , \bar \f)}S_0
+ \Big\{     \int {\rm d}^{4} x \rd^2 \q\,
{\cal E} \, S_0^3   W(\f)
+ {\rm c.c.} \Big\}~.~~~
\eea
This corresponds to 
old minimal  supergravity, 
with $\f$ being  matter chiral superfields taking their values in a K\"ahler manifold,
with $K(\f , \bar \f)$ its K\"ahler potential. 
In this case $\U$ in the generalised FI term in \eqref{CompleteAction}
should be $\U = \bar S_0  \re^{-\frac{1}{3} K(\f , \bar \f)}S_0 $, as was pointed out 
in \cite{ACIK2,AKK}, 
and then our model can be written in a form similar to that with  the standard FI term \cite{SW},
\bea
-3 \U -2 \x \U {\mathbb V}^{(\cG)} =-3  \bar S_0  \,
\re^{-\frac{1}{3} K(\f , \bar \f) +\frac 23 \x {\mathbb V}^{(\cG)}}S_0~.
\eea

In general, 
the vector multiplet part of the action  \eqref{CompleteAction}
is highly nonlinear. 
However its functional form drastically simplifies if
the ordinary gauge field contained in $V$ is a flat connection.
Then the gauge freedom \eqref{gaugefreedom}
allows one to choose a gauge in which  $V$ is 
a nilpotent superfield constrained by 
\bea
{V}{V}=0~, 
\qquad
{V} \cD_A \cD_B {V}=0~,
\qquad
{ V} \cD_A \cD_B \cD_C {V} =0~.
\label{14}
\eea
In this gauge
it can be shown  (see also \cite{KMcAT-M}) that 
\bea
V= - 4 \frac{\cW^2 \bar \cW^2}{(\cD \cW)^3} \quad \implies \quad
V \cD^2 \cW^2 = - V(\cD \cW)^2 \quad \implies \quad
 {\mathbb V}^{(\cG)} = \cG(1,1)V~.
 \label{1.11}
 \eea
These relations imply that the $V$-dependent part of the action 
\eqref{CompleteAction} becomes
\bea
S[V] -2\x {\mathbb J}^{(\cG)}_{\rm FI} [V; \U] 
~\longrightarrow  ~\frac{h}{2}  \int \rd^4 x \rd^2 \q   \, \cE\,   
\cW^2 -2\x g \int \rd^4 x \rd^2 \q  \rd^2 \bar{\q} \, E\,
\U V~, 
\label{Goldstino}
\eea
where we have denoted  $h:=1 + \hf \cH (1, 1)$ and $ g := \cG(1,1)$.
Under mild conditions 
\bea
\cH(1,1) > - 2~, \qquad \cG(1,1) \neq 0~, 
\label{2.15}
\eea
the functional \eqref{Goldstino} coincides (modulo an overall numerical factor)
with the  Goldstino multiplet action, $S_{\rm Goldstino}$, 
which was proposed in \cite{KMcAT-M} and is given by 
\bea
S_{\rm Goldstino}&=& \frac{1}{2}  \int \rd^4 x \rd^2 \q   \, \cE\,   
\cW^2 -2\x  \int \rd^4 x \rd^2 \q  \rd^2 \bar{\q} \, E\,
\U V~, 
\eea
where the Goldstino superfield $V$ obeys the constraints \eqref{1.11}
and is subject to the condition that $\cD \cW $ be nowhere vanishing.
It should be noted that the condition $h>0$
is equivalent to the requirement that the kinetic term of the Goldstino field has
 the correct sign.
  At the component level, the action \eqref{Goldstino}
is highly nonlinear due to the nilpotency constraints  \eqref{14}.
However, the functional form of  \eqref{Goldstino} is universal, 
unlike the complete vector multiplet action in \eqref{CompleteAction}, which is a manifestation of the fact that the Volkov-Akulov action \cite{VA} 
is universal \cite{IK}.\footnote{The constraints 
\eqref{14} are invariant under local rescalings $V \to \re^{\r} V$, with the parameter $\r$ being an arbitrary real scalar superfield. Requiring the action \eqref{Goldstino} to be 
stationary under such rescalings gives the constraint 
$f \U V = - \hf V \cD^\a \cW_\a = \frac{1}{8} V \cD^\a (\bar \cD^2 -4R ) \cD_\a V$, where $f=\x g/h$. 
In conjunction with \eqref{14},  this constraint defines
the irreducible Goldstino multiplet introduced in \cite{BHKMS}.}
We have thus shown that the supergravity-matter system  \eqref{CompleteAction}
describes spontaneously broken local supersymmetry under the condition 
\eqref{2.15}.

To arrive at  \eqref{Goldstino}, we have made use of the gauge fixing \eqref{14}, 
which exists only if the component gauge field is a flat connection and 
 which expresses the fact that the gauge field is switched off.
 This is actually possible to avoid by: (i) making 
 any assumption about the component gauge field; 
  and (ii) imposing any gauge condition. 
In general, the unconstrained gauge prepotential $V$ may be split 
in two superfields, one of which 
contains  the U(1) gauge field and all  purely gauge degrees of freedom, 
while the other contains the remaining physical component fields.
The point is that the nilpotency conditions \eqref{5} allow us to interpret 
${\mathbb V}^{(\cG)}$
as a Goldstino superfield of the type proposed in \cite{KMcAT-M} provided
 its $\sD$-field is nowhere vanishing, which means that $\cD^2 {\cW}^2$ is nowhere vanishing, 
 in addition to the condition $\cD \cW\neq 0$ imposed earlier. 
Then ${\mathbb V}^{(\cG)}$ contains only two independent component fields, 
the Goldstino and $\sD$-field. We then can introduce a  new parametrisation 
for the gauge prepotential given by 
\bea
V = A^{(\cG)} + {\mathbb V}^{(\cG)}~.
\eea
 It is $A^{(\cG)} $ which varies under the gauge transformation \eqref{gaugefreedom},
$\d_\l A^{(\cG)}  = \l + \bar \l$, while ${\mathbb V}^{(\cG)}$ is gauge 
invariant by construction.
Modulo purely gauge  degrees of freedom, $A^{(\cG)} $ contains a single  independent 
physical field, the gauge one-form.

Our discussion shows that a flat-superspace  limit of the vector multiplet 
action $S[V] -2\x {\mathbb J}^{(\cG)}_{\rm FI} [V; \U] $ in \eqref{CompleteAction}
gives a consistent model for spontaneously broken rigid supersymmetry 
under the conditions \eqref{2.15}.


\section{The component structure} 

It is important to analyse the bosonic Lagrangian of the model 
\eqref{CompleteAction} in the vector multiplet sector. For this purpose 
we  introduce  gauge-invariant component fields of the vector multiplet following \cite{KMcC}
\bea
\cW_{\a}\arrowvert = \c_{\a}~,\qquad
-\frac{1}{2}\cD^{\a} \cW_{\a}\arrowvert=\sD~,
\qquad
\cD_{(\a} \cW_{\b)}\arrowvert 
=2 {\rm i} {\hat F}_{\a\b}
&=& {\rm i} (\s^{ab})_{\a\b}{\hat F}_{ab}~,
\label{19}
\eea
where the bar-projection  $U|$ of a superfield $U$ means switching off the superspace Grassmann variables, and 
\bea
\label{eq:F_ab defn}
{\hat F}_{ab} &=& F_{ab} -
\frac{1}{2}(\J_{a}\s_{b}{\bar \c} + 
\c\s_{b}{\bar \J}_{a}) +
\frac{1}{2}(\J_{b}\s_{a}{\bar \c} + 
\c\s_{a}{\bar \J}_{b})~,
\non\\
F_{ab} &=& { \nabla}_{a}V_{b} 
- { \nabla}_{b}V_{a} 
- {\cT_{ab}}^{c}V_{c}~,
\eea
with $V_a= e_a{}^m (x) \,V_m (x)$  
the  gauge  one-form, and $\J_a{}^\b$ the gravitino. 
Here ${ \nabla}_a$ denotes a Lorentz-covariant derivative with torsion,
\bea
\label{eq:covariant derivative algebra}
\left[{ \nabla}_{a}, { \nabla}_{b}\right] = {\cT_{ab}}^{c} { \nabla}_{c}
+ \frac{1}{2}\,\cR_{abcd} M^{cd}~,
\eea
where  $\cT_{abc}$ and $\cR_{abcd}$ are the torsion and curvature tensors,
respectively. The former  is determined by  the gravitino via
$
\cT_{abc} = -\frac{\rm i}{2}(\J_{a}\s_{c}{\bar \J}_{b}
- \J_{b}\s_{c}{\bar \J}_{a}).
$
Making use of the above relations leads to
\bea
-\frac{1}{4} \cD^2 {W}^2| = \sD^2 - 2 F^2 +
\text{fermionic terms}~, \qquad F^2: =F^{\a\b} F_{\a\b}~.
\eea
We conclude that the electromagnetic field should be weak enough 
to satisfy $\sD^2 - 2 F^2  \neq 0$,
in addition to the condition $\sD \neq 0$ discussed above.
Direct calculations yield the component bosonic Lagrangian 
\bea
\cL (F_{ab} , \sD)= -\hf (F^2 +\bar F^2) &+&\hf \sD^2\left\{
1+ \hf  \cH \Big(1 -  \frac{2F^2}{\sD^2}\, ,\, 1 -  \frac{2\bar F^2}{\sD^2} \Big) 
\Big| 1 -  \frac{2F^2}{\sD^2} \Big|^2
 \right\} \non \\
&-&\x  \sD\, 
\cG \Big(1 -  \frac{2F^2}{\sD^2}\, ,\, 1 -  \frac{2\bar F^2}{\sD^2} \Big) 
\Big| 1 -  \frac{2F^2}{\sD^2} \Big|^2 \U|~.
\label{23}
\eea
In order for the supergravity action in 
\eqref{CompleteAction}
to give  the correct Einstein-Hilbert gravitational Lagrangian at
the component level, the super-Weyl gauge $\U|=1$ should be imposed, 
see \cite{KMcC,KU} for the technical details.

Let us briefly discuss the structure of the Lagrangian \eqref{23}.
It is seen that the case \cite{CFTVP} 
\bea
\cG (z, \bar z) = \frac{g} {z \bar z}
\label{3.7}
\eea
is special since the last term in \eqref{23}
becomes linear in $\sD$ and independent of the field strength.
Another special case corresponds to 
\bea
\cH (z, \bar z) = \frac{2h-1} {z \bar z}~,
\label{3.8}
\eea
since the second term in \eqref{23} becomes quadratic in $\sD$ and independent of $F$. 
The following table summarises the differences between the standard and generalised
 FI terms:
\bea
\begin{array}{|l|l|}
\hline
\phantom{\bigg|}\mbox{~~Standard FI term \cite{FI}} ~ {} & ~~ {J}_{\rm FI} = \hf \sD~ \\
  \hline
\phantom{\bigg|}\mbox{~~Special  FI-type term  \cite{CFTVP}} ~~&
~~ {\mathbb J}^{[-1]}_{\rm FI} = \hf \sD+ O(\c)   \\
\hline
\phantom{\bigg|} \mbox{~~General   FI-type term \cite{K18,K19}} ~~&
~~
{\mathbb J}^{(\cG)}_{\rm FI} = \hf \sD+ O(\c , F)~~ \\
\hline
\end{array}
\qquad {}
\eea
Making use of the local supersymmetry allows one to impose the unitary gauge $\c_\a=0$,
which is why little is lost in restricting our analysis to the bosonic sector \eqref{23}.
Imposing the gauge $\c_\a=0$ reduces   ${\mathbb J}^{[-1]}_{\rm FI}$ to 
the standard FI term, $ \hf \sD$.

As follows from \eqref{23}, 
the auxiliary field $\sD$ may be integrated out  using its equation of motion 
\bea
\frac{\pa}{\pa \sD} \cL (F_{ab} , \sD)=0~,
\eea
resulting in a model for nonlinear electrodynamics, ${\mathfrak L} (F_{ab})$.
The special feature of the choice \eqref{3.7} and \eqref{3.8}
is that ${\mathfrak L} (F_{ab})$ coincides with Maxwell's Lagrangian.

\section{Generalised FI terms as quantum corrections}

In section 1, it was pointed out that all healthy mechanisms to generate a cosmological 
constant should explain both (i) its positivity and (ii) its small value. So far we have only discussed point (i). Here  a comment will be made regarding point (ii).

It is known that the standard FI term in rigid supersymmetric gauge theories
does not receive any quantum corrections \cite{FNPRS,GrisaruSiegel}.
In other words, there is no way to generate an FI term quantum mechanically. 
It appears that  generalised FI terms may occur as pure quantum corrections. 
This hope is supported by the fact that the nonlinear term in the superconformal action \eqref{action} 
has a functional form typical for quantum corrections to low-energy effective actions 
in SYM theories, see e.g. \cite{McAG,PB}. However, the important difference between the functionals \eqref{action} and \eqref{generalised} is that the former is even  with respect to the replacement $V\to -V$, while  the latter is odd (similar to the  anomalous  effective action in supersymmetric gauge theories). Explicit calculations will be reported elsewhere.

${}$\\

\noindent
{\bf Acknowledgements:} 
It is my pleasure to thank the organisers of the Workshop SQS'2019
for their warm hospitality in Yerevan. I am grateful to Darren Grasso and Ulf Lindstr\"om for comments on the manuscript.
The research presented in this  work was supported in part by the Australian 
Research Council, projects DP160103633 \& DP200101944.


\begin{footnotesize}

\end{footnotesize}

\end{document}